\documentclass[useAMS,usenatbib]{mn2e}

\usepackage{graphicx}
\usepackage{times}
\newcommand{\clsshort}{XLSSC~046}
\newcommand{\cls}{XLSSJ022303.0-043622}
\newcommand{\eg}{{\it e.g.~}}
\newcommand{\ie}{{\it i.e.~}}
\newcommand{\ergps}{erg s$^{-1}$}
\newcommand{\ergpspcms}{erg cm$^{-2}$s$^{-1}$}

\title[XMM-LSS discovery of a $z=1.22$ galaxy cluster]{XMM-LSS
discovery of a $z=1.22$ galaxy cluster} \author[M.N. Bremer et al.
]{M.N. Bremer$^1$\thanks{E-mail: m.bremer@bristol.ac.uk},
I. Valtchanov$^{2,3}$, J. Willis$^4$, B. Altieri$^3$, S. Andreon$^5$,\newauthor  P.A
Duc$^6$, F. Fang$^{7,8}$, C. Jean$^9$, C. Lonsdale$^8$, F. Pacaud$^6$,
M. Pierre$^6$,\newauthor J.A. Surace$^{7,8}$, D.L. Scupe$^{7,8}$, I. Waddington$^{10}$ \\
$^1$ H.H. Wills Physics
Laboratory, University of Bristol, Tyndall Avenue, Bristol BS8 1TL,
UK.\\ $^{2}$Blackett Laboratory, Astrophysics Group, Imperial College,
London SW7 2BW, UK.\\
$^{3}$ESA, Villafranca del Castillo, Spain.\\ 
 $^{4}$Department of Physics and Astronomy, University of Victoria, Elliot Building, 3800 Finnerty Road, Victoria, BC, V8P 1A1 Canada.\\ 
$^{5}$INAF   Osservatorio Astronomico di Brera, Milan, Italy.\\
$^{6}$CEA/DSM/DAPNIA, Service d'Astrophysique, Saclay, F-91191
Gif sur Yvette, France.\\
$^7$Spitzer Science 
Center, California Institute of Technology, MS 220-6, Pasadena, CA, 
91125, USA.\\
 $^{8}$Infrared Processing and Analysis center, California Institute 
of Technology, MS 100-22, Pasadena, CA, 91125.\\
Center, Caltech, USA. \\ 
$^{9}$Institut d Astrophysique et de G\'eophysique, ULg, All\'ee du 6 AoÆut 17, B5C, 4000 Sart Tilman (Li\`ege), Belgium.\\
$^{10}$Astronomy Centre, University of Sussex, Falmer, Brighton BN1 9QH, UK.}
\begin{document}

\date{Accepted ??. Received \today; in original  form \today}

\pagerange{\pageref{firstpage}--\pageref{lastpage}} \pubyear{2005}

\maketitle

\label{firstpage}

\begin{abstract}
We present details of the discovery of \cls, a $z=1.2$ cluster of
galaxies. This cluster was identified from its X-ray properties and
selected as a $z>1$ candidate from its optical/near-IR characteristics
in the XMM Large-Scale Structure Survey (XMM-LSS). It is the most
distant system discovered in the survey to date. We present
ground-based optical and near IR observations of the system carried
out as part of the XMM-LSS survey. The cluster has a bolometric X-ray
luminosity of $ 1.1\pm0.7 \times 10^{44}$ \ergps, fainter than most other
known $z>1$ X-ray selected clusters.  In the optical it has a
remarkably compact core, with at least a dozen galaxies inside a 125
kpc radius circle centred on the X-ray position. Most of the galaxies
within the core, and those spectroscopically confirmed to be cluster
members, have stellar masses similar to those of massive cluster
galaxies at low redshift. They have colours comparable to those of
galaxies in other $z>1$ clusters, consistent with showing little sign
of strong ongoing star formation.  The bulk of the star formation
within the galaxies appears to have ceased at least 1.5 Gyr before the
observed epoch.  Our results are consistent with massive cluster
galaxies forming at $z>1$ and passively evolving thereafter. We
also show that the system is straightforwardly identified in
Spitzer/IRAC $3.6\micron$ and $4.5\micron$ data obtained by the SWIRE
survey emphasising the power and utility of joint XMM and Spitzer
searches for the most distant clusters.

\end{abstract}

\begin{keywords}
galaxies -- clusters; large-scale structure
\end{keywords}

\section{Introduction}

Determining the properties of clusters and their constituent galaxies
at $z>1$ is central to understanding how such systems evolve with
cosmic time. Distant clusters are identifiable using multiple
techniques across a number of wavebands, from optical and near IR
imaging surveys that identify emission from their galaxies to X-ray
surveys that identify emission from the gaseous Intra Cluster Medium
(ICM). As the X-ray identification of a cluster is straightforwardly
related to important physical parameters such as the mass of a system,
X-ray selected samples have been the most heavily studied to date.
Although X-ray work had been carried out beforehand (notably with the
EMSS, \citealt{gioia90}), it was work using the ROSAT satellite which
allowed the routine discovery and study of clusters out to $z\sim
0.8$, (\eg \citealt{boh04, ros02, vik98,mul03}). Confirmed $z>1$
clusters remain rare beasts despite well over a decade of searches,
with only a handful known from the ROSAT era (\eg \citealt{vik98,
ros99, stanford02, blak03}). The potential of XMM-Newton to detect clusters
out to $z\sim 2$ was demonstrated as early as a year and a half before
XMM launched \citep{val00}. Recently \cite{mul05} and \cite{stan06} have shown that
clusters out to $z\sim 1.5$ are serendipitously detectable in
moderately deep XMM exposures.

Work to date on X-ray selected clusters has shown that there is
little evidence for evolution in their comoving space density (except
for the most luminous systems) or $L_X-T_X$ relation out to $z\sim 1$
\citep{ros02}. Similarly, there is little evidence for anything other
than passive evolution in the stellar populations of the massive
galaxies in the cores of these clusters (\eg \citealt{blak03}) over
the currently observed redshift range. Strong evolution
of the ICM and galaxy populations in clusters seems to have occurred
only at $z>1$. Increasing the number of known and well-studied $z>1$
clusters is therefore key to identifying this epoch and understanding
the physical processes that drove the evolution of clusters of
galaxies.

We are carrying out the XMM Large-Scale Structure Survey (XMM-LSS,
\citealt{pie04}) in order to study the large-scale distribution of
matter in the Universe as traced through X-ray emission from clusters
of galaxies and active galactic nuclei (AGN). One of the survey goals
is to search for clusters of galaxies out to $z>1$ to luminosities
$L_X([0.5-2.0]~\rm{keV}) > 5 \times 10^{43}$ \ergps (\ie a flux limit of
$f_X([0.5-2.0]~\rm{keV}) \sim 10^{-14}$\ergpspcms). Here we present details
of the discovery of a cluster with the highest
spectroscopically-confimed redshift, $z=1.22$, so far found in the
XMM-LSS. Some initial photometric data on this cluster was presented
in \cite{andreon05}.

This source was discovered in a survey of a relatively large
contiguous area of sky ($\sim 5$ deg$^2$), specifically designed to detect
clusters to $z>1$. This has the advantage that existing
multi-wavelength datasets covering the same area can be used in
combination to identify clusters from among the faint, extended X-ray
sources in the XMM data, whereas for sources serendipitously detected
in archived non-contiguous X-ray data, the multi-wavelength data have
to be obtained after the identification of X-ray candidates. The
disadvantage of the survey approach is that the X-ray exposure times
are necessarily limited in order to cover sufficient area of sky (in
the case of the XMM-LSS to typically 10-20 ksec), whereas the
serendipitous approach can make use of far deeper exposures.

In the following we present the X-ray, optical and near-IR data for
\cls, including Spitzer/IRAC 3.6\micron\ and 4.5\micron\ band imaging
from the SWIRE survey. We discuss the cluster properties, briefly
comparing them to those of other known $z>1$ clusters and demonstrate that a
combined XMM and Spitzer survey is an efficient way of identifying
further high redshift clusters. The cluster is also catalogued as
\clsshort\footnote{See
  http://vizier.u-strasbg.fr/cgi-bin/Dic-Simbad?XLSSC}. We use
$\Lambda$CDM cosmology ($\Omega_m = 0.3,\ \Omega_{\Lambda}=0.7,\ H_0 =
70$ km s$^{-1}$ Mpc$^{-1}$).

\section{Observations}

\subsection{X-ray}

\begin{figure}
\includegraphics[width=8cm]{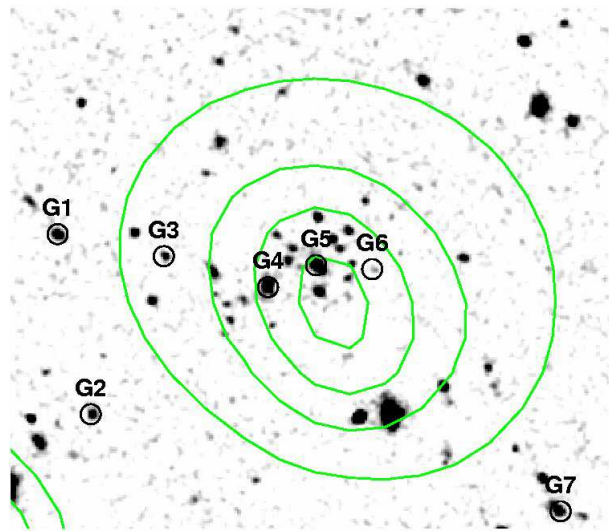}
\caption{Sofi $Ks-$band image of the central $\sim$ 1.5 by 1.5
arcmin$^2$ region of \cls \ smoothed with a gaussian of FWHM
$0.9\arcsec$. Overlayed as contours is the wavelet filtered XMM
[0.5-2] keV image. The lowest contour has a diameter of $\sim
1\arcmin$, contours increment logarithmically in surface brightness
with and integrated total flux of 58 counts. Galaxies with measured
redshifts are labelled as in table 1. North is up, East to the left.}
\label{fig:cls}
\end{figure}

The XMM X-ray data for the field of this cluster (XMM exposure ID
0109520601) were obtained as part of the first set of XMM-LSS
pointings during GTO and AO-1 time. The resulting event list for the
field was filtered for high background periods following \citet{pra02}
with a resulting exposure time of 51.6 ks across the three XMM
detectors (MOS1: 22.5 ks; MOS2: 22.7 ks; pn: 16.4 ks).  Less than 1\%
of the MOS and $\sim 4$\% of the pn exposure was affected by periods
of high background.  The ``cleaned'' event list was then used to
create images in different energy bands for the pn and MOS detectors.
Using the procedure in \citet{val01}, potential candidate clusters
were selected as extended X-ray sources in the [0.5-2] keV band,
taking into account the effect of vignetting.

\begin{figure*}
\includegraphics[width=15.5cm]{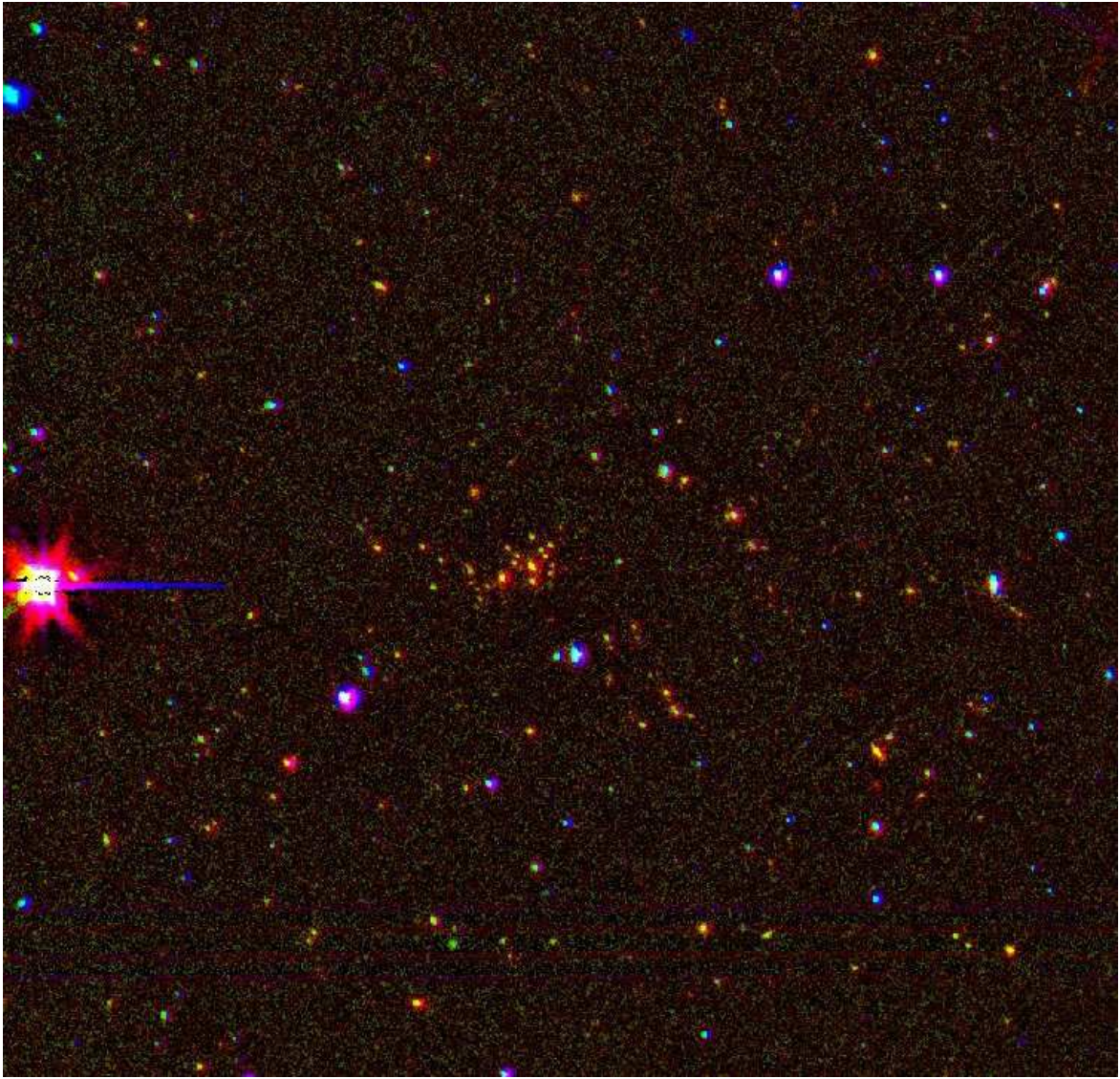}
\caption{Three colour composite image of the $4.3\arcmin$ by
$4.4\arcmin$ field centred on \cls, made from VLT/FORS2 $Z-$band
(blue), NTT/SOFI $Ks-$band (green) and Spitzer/IRAC 3.6$\micron$ band
images (red).North is up, East is to the left. Although there other
objects in the field with similar colours and magnitudes as those in
the cluster, the surface density of such galaxies at the cluster
centre is far higher than in the rest of the field. The cluster
galaxies are clearly red, indicating their strength at 3.6
\micron. While the galaxies are individually detectable in the 
optical data, their red colours make them stand out as an overdensity
in the $Ks$ and IRAC bands.
\label{tci}}
\end{figure*}

The cluster was detected as an extended X-ray source at an off-axis
angle of $8.8\arcmin$ using an early version of the XMM-LSS analysis
pipeline. At this angle an incoming X-ray photon sees 60 per cent of
the telescope area due to vignetting. Placing a $34\arcsec$ \ radius
aperture over the source, and correcting for background counts gave 40
counts for the object in the [0.3-10] keV band of the MOS detectors and
45 in the PN detector.  In the intervening time between the detection
of the X-ray source and now, we have evolved our X-ray analysis based
on the experience gained from the early XMM-LSS data. In section 3 we
discuss the X-ray luminosity and temperature of the cluster as derived
from our latest analysis methods (discussed in detail in
\citealt{willis05} and \citealt{d1}). Here we note that the source is
detected and classified by the latest version of the XMM-LSS pipeline
as a ``C2'' source (see \citealt{pacaud05} for details).

\subsection{Optical and Near-IR data}

\begin{figure}
\includegraphics[width=8cm]{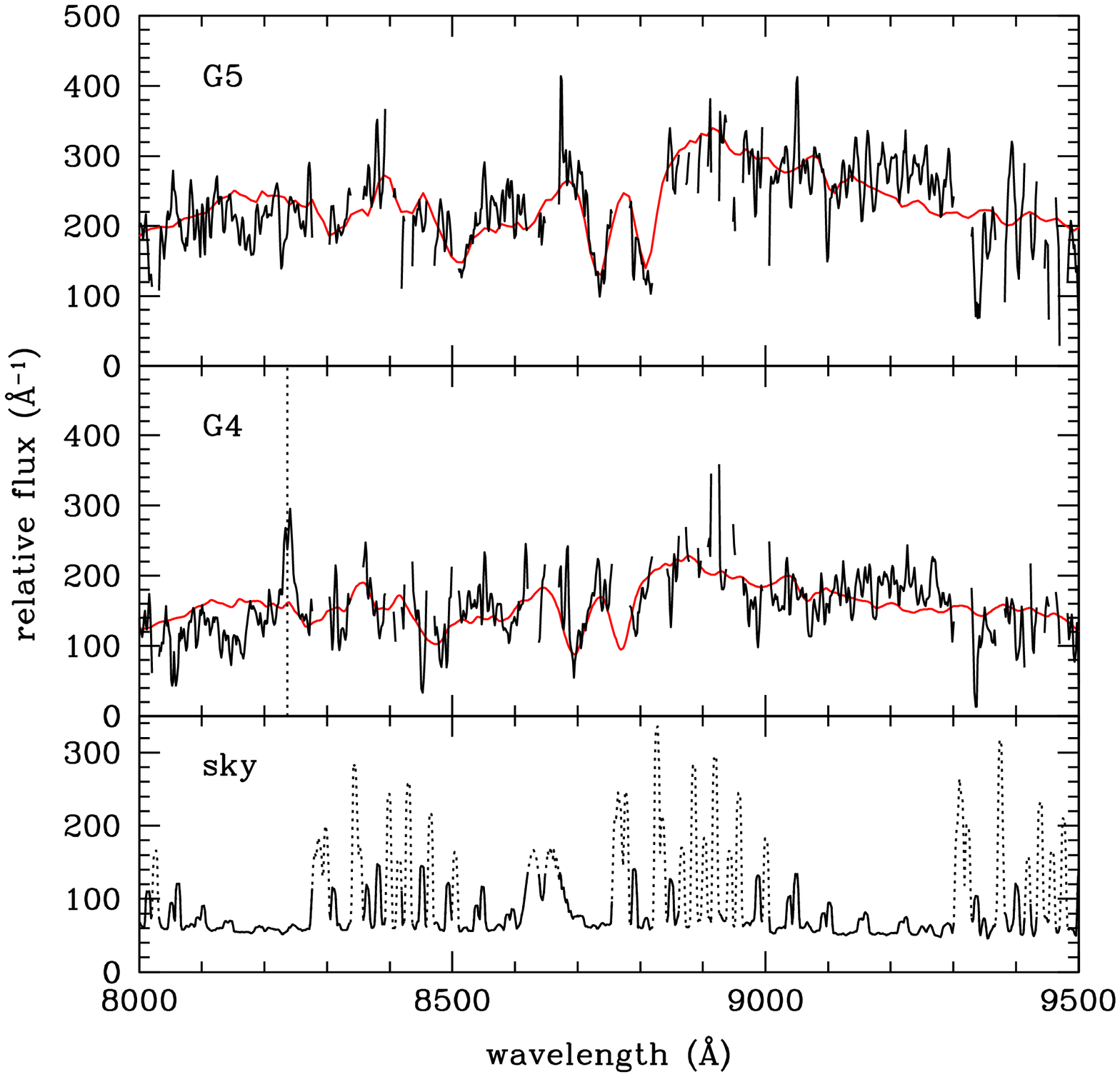}
\caption{The top and middle panels show spectra of the two brightest
galaxies in the cluster (solid line). Wavelength regions associated
with bright sky emission lines have been excised from the spectra --
which have subsequently been resampled to match the spectrograph
resolution. Total exposure time for each spectrum is 2 hours. The
smooth line in each panel shows a low redshift elliptical galaxy
template \citep{kin96} that has been redshifted and scaled to match
the data on wavelength scales $> 500$\AA. The vertical dotted line in
the middle panel indicates the observed frame location of
[OII]~$\lambda$3727 emission. The lower panel displays the sky
spectrum which is shown as a dotted line in wavelength regions excised
from the spectra plotted above.  The spectra are corrected for
relative flux calibration, although this correction is uncertain
longward of 9000 \AA. }
\label{fig:spec}
\end{figure}

On their own, the observed X-ray properties of the source are compatible
with it being a nearby galaxy, a moderate redshift group or a high
redshift cluster. Additional multi-wavelength data is required to
determine its nature.  The initial XMM-LSS sky area was imaged in
several optical bands with the CFH12k camera on CFHT
\citep{vvds,mac03}. The lack of a clear identification for \clsshort \ in
this relatively shallow optical data made the object a strong distant cluster
candidate. In particular, there was no clear overdensity of galaxies
(or an individual low redshift galaxy) at the X-ray position down to
$I_{AB}=22$, indicating that the X-ray source was a potential $z>1$
cluster. Subsequently, the field of the cluster was observed during
the Canada-France-Hawaii Telescope Legacy Survey (CFHTLS) using Megacam on CFHT
\citep{boul03}. The $i'$ and $z'$ photometry used in this paper is
derived from the CFHTLS data.

As a result of the lack of a clear optical ID in the CFH12K data, the
source was targetted as part of an ESO NTT/SOFI observing run to
follow up potential distant clusters (programme ID 70.A-0733A) on 2002
November 19 and 20. The cluster was observed for 40 minutes in each of
$J$ and $Ks$ in photometric conditions at an airmass below 1.5, with
observations divided into multiple, dithered exposures.  These data
were calibrated by observations of photometric standards sj9105 and
sj9106 \citep{persson98}. Five 2MASS-catalogued objects were detected
in the frames, the photometry for each had a typical uncertainty of
0.1 magnitude in the 2MASS catalogue; our calibration matches that of
2MASS within these uncertainties.  All $J$ and $Ks$ magnitudes
presented in this paper have been converted to AB magnitudes assuming
$J_{AB}=J_{Vega}+0.9$ and $Ks_{AB}=Ks_{Vega}+1.9$ for consistency with
the CFHTLS and IRAC magnitudes, which are measured in this system.

A high surface density of faint galaxies ($19.15<Ks<21.65$, or
$17.25<Ks<19.75$ in Vega) -- fifteen in a $30\arcsec$ diameter circle
(a projected scale of 250 kpc at $z=1.22$) -- was identified in both
near-IR bands at the position of the X-ray detection (see figures
\ref{fig:cls} and \ref{tci}). A circle of twice that diameter includes
only a further eight objects in the same magnitude range, indicating
the compactness of the galaxy distribution. All of this is compatible
with a compact cluster of galaxies at $z>1$. 

Spectroscopic observations of the cluster were obtained using the
ESO/VLT FORS2 spectrograph on 2004 December 14 and 15. Multi-slit
spectroscopic data were obtained using the 600z$+$23 grism and
OG590$+$32 order-sorting filter. The slit width on each MXU mask was
typically $1.2\arcsec$ and resulted in a spectral resolution of 8\AA\
over the wavelength interval 7400 to 10000\AA. The cluster was
observed with two slit masks oriented at 90 degrees to each other in
order to maximise the sampling of candidate cluster members. Each mask
was observed for $4 \times 30$~minute exposures at an airmass below
1.5 and typical atmospheric seeing of $0.8\arcsec$. Any given object
was observed through one slit mask giving total exposure times of two
hours per galaxy. Spectral data were
processed employing standard reduction techniques within {\tt IRAF}.
Zero level, flat--field and cosmic ray corrections were applied to all
data prior to the identification, sky subtraction and extraction of
individual spectra employing the {\tt apextract} package. The
dispersion solution for each extracted spectrum was determined
employing HeNeAr lamp exposures with a typical residual scatter of
$0.5-0.8$\AA\ and the final spectra were resampled to a linear
wavelength scale. A spectrophotometric standard star from the atlas of
\cite{hamuy92} was observed during each night and was employed to
correct for the relative instrumental efficiency as a function of
wavelength.

Reliable redshift values consistent with a cluster redshift of $z=1.22
\pm 0.01$ were estimated for seven galaxy spectra (see table
1). Initial redshift estimates obtained from a visual inspection of
prominent absorption and emission features in individual spectra were
refined via cross--correlation with a representative early--type
galaxy template \citep{kin96} employing the {\tt IRAF} routine {\tt
xcsao} \citep{tonry79}. Examples of extracted spectra are displayed in
figure \ref{fig:spec}. The location of all galaxies confirmed at the
cluster redshift are displayed in figure \ref{fig:cls}.

\subsection {SWIRE Spitzer observations}

The field of the cluster was imaged by the IRAC camera on the Spitzer
satellite as part of the SWIRE \citep{lon03} survey of the XMM-LSS sky
area. Images of the field at 3.6 and $4.5\micron$ -- reduced using the
standard SWIRE processing \citep{lon04} -- clearly show the cluster
(the 3.6 \micron \ image is used as the red channel in figure
\ref{tci}). The high surface density of the galaxies at the very centre of
the cluster, coupled with the relatively poor spatial resolution of
IRAC at 3.6 and $4.5\micron$ leads to high crowding at the cluster
centre and so to difficult photometry in this region. The
rest of the field is uncrowded making for straightforward
photometry. The field was also imaged at 24 \micron \ by the MIPS
camera. No cluster members were detected to a limit of 0.15mJy
(fainter than 18.6 in AB). Unfortunately this limit hardly constrains
the star formation activity in the galaxies. At the cluster redshift
it translates to a luminosity of less than $L_{IR}<10^{12}$L$_\odot$,
so no cluster galaxies host obscured ULIRG-like starbursts.

\subsection{Photometry}

In order to carry out photometry on our ground-based optical and
near-IR data sets, we used Sextractor \citep{bert96} to identify
individual objects and determine their magnitudes. Results for the six
spectroscopically-confirmed cluster galaxies brighter than $K_s<21.65$
are given in table 1. We selected the catalogue in the $Ks-$band,
using the $Ks$ SOFI image as the finding image for other bands.

\begin{table*}

\begin{tabular}{llllcccccc}
ID& Redshift &RA(2000)&Dec(2000)& $Ks$&$i'-Ks$&$i'-z'$&$J-Ks$&$Ks-3.6$&$3.6-4.5$ \\
 \hline
G1& 1.219& 35.7739 & -4.6030 & $ 20.31\pm0.05$ & $  3.07\pm0.10$ & $  0.90\pm0.06$ & $  0.94\pm0.09$ & $  0.66\pm0.12$ & $ -0.25\pm0.11$\\
G2& 1.215& 35.7725 & -4.6102 & $ 20.80\pm0.08$ & $  2.75\pm0.12$ & $  0.79\pm0.07$ & $  0.98\pm0.14$ & $  0.74\pm0.14$ & $ -0.26\pm0.11$\\
G3& 1.215& 35.7696 & -4.6039 & $ 21.10\pm0.10$ & $  3.04\pm0.14$ & $  0.64\pm0.13$ & $  0.68\pm0.15$ & $  0.69\pm0.14$ & $ -0.26\pm0.12$\\
G4&1.210& 35.7655 & -4.6051 & $ 19.78\pm0.04$ & $  3.01\pm0.08$ & $  0.72\pm0.06$ & $  1.13\pm0.07$ & $  0.69\pm0.17$ & $ -0.17\pm0.10$\\
G5&1.221& 35.7635 & -4.6043 & $ 19.62\pm0.04$ & $  3.21\pm0.07$ & $  0.87\pm0.06$ & $  0.97\pm0.05$ & $  0.58\pm0.19$ & $ -0.19\pm0.10$\\

G7&1.210&  35.7539 & -4.6140 & $ 20.31\pm0.05$ & $  2.56\pm0.10$ & $  0.88\pm0.06$ & $  0.80\pm0.08$ & $  0.55\pm0.20$ & $ -0.28\pm0.11$\\

\hline

\end{tabular}
\caption{$i',z'$ (from CFHTLS),$J,Ks$,(SOFI) and 3.6\micron,
  4.5\micron (IRAC) photometry for six of the
  spectroscopically-confirmed cluster members with $Ks<21.65$ (all
  colours and magnitudes in AB). Galaxy G6 ($z=1.224$) was too faint
  to be included in the $Ks-$selected sample.  }

\end{table*}

SEXTRACTOR was used in dual mode using the
$Ks$ image to identify objects and to define the apertures that were then
applied to the other optical and near-IR images.  The $Ks-$band
detected number counts peaked between $Ks=21.4$ and 21.65 and dropped
to 50 per cent of the peak between $Ks=22.15$ and 22.4. We carried out
the photometry in two ways: firstly by using 2 arcsec diameter fixed
apertures with appropriate aperture corrections in each band and
secondly by using the corrected isophotal ``AUTO'' magnitude (after
Kron 1980) having previously smoothed all the data to 0.8 arcsec FWHM,
matching the worst seeing image (the $I-$band). Both methods gave
comparable results, in the following we report the 2 arcsec aperture
magnitudes as the smoothing process complicates the determination of
uncertainties for the corrected isophotal magnitudes.

Using the $Ks$ frame as the finding image was straightforward for the
$J-$band photometry because the data shared the same plate scale and
astrometric solution. For the Megacam data, we resampled the $Ks$
image to the Megacam plate scale to act as the finding image. The same
objects and apertures were re-identified in the resampled image as in
the original $Ks$ frame and applied to the $i'$ and $z'$ data. The
resampling meant that the SEXTRACTOR coordinates (and therefore the
aperture centre) for each object changed slightly, but by no more than
0.15 arcsec (and usually $<0.1$ arcsec, less than half a pixel) in the
case of the cluster members listed in table 1.  As a check on the
effect of resampling on the determination of magnitudes, we compared
the results of photometry on the original and resampled $Ks$
image. For the galaxies of comparable magnitudes to those in listed in
table 1, this introduced a scatter of no more than 0.05
magnitudes. Additionally we shifted the resampled image by 0.5 pixels
in both RA and DEC to see what effect that had on the determination of
the $i'$ and $z'-$band magnitudes. For galaxies of similar magnitudes
to those in table 1, a typical scatter of 0.025 magnitudes was induced
by this process. To account for these the uncertainties quoted in
table 1 for $i'-Ks$ and $z'-Ks$ include the statistical uncertainties
from SEXTRACTOR and a 0.05 magnitude systematic uncertainty added in
quadrature.

For the IRAC photometry, we used the catalogue made by the SWIRE
collaboration from the SWIRE data release 2 (described in
\citealt{surace05}) as this provides aperture photometry with
appropriate aperture corrections. Following the recommendations in
Surace etal we used apertures of diameter 3.8 arcsec (twice the
instrumental FWHM) which provide the most accurate photometry and have
had aperture corrections for unresolved sources (like the cluster
galaxies) already applied. The IRAC aperture centroids matched those
in the K-band to within 0.2 arcsec for all objects in table 1 except
for G3 which matched within 0.4 arcsec.  For galaxies G4,5 and 7, the
3.8 arcsec apertures included emission from fainter close companion
galaxies. To correct for this, we determined the difference in
$Ks$-band flux between 2 and 3.8 arcsec apertures centred on these
objects, assumed that the companions had similar $K-3.6\micron$
colours to the brighter galaxies and adjusted this colour for each of
the objects accordingly (a correction of between 0.3 and 0.37
magnitudes in each case). The assumption of similar colours for the
companion objects is reasonable as they are likely cluster members,
nevertheless this is a source of uncertainty. Assuming a potential
colour differential of $-0.3<\Delta(K-3.6)<0.3$ between the main galaxy and a
companion contributing 30 per cent of the flux in an aperture, this
would give an uncertainty of just les than 0.1 magnitudes in the colur
determined in this way. This is reflected in the uncertainties quoted
 for the $K-3.6\micron$ colours of these objects in table 1. The
absolute accuracy quoted by the SPITZER Science Centre for IRAC is
10\% in flux density. The $K-3.6\micron$ and $3.6-4.5\micron$
uncertainties listed in table 1 include this and the statistical
uncertainties combined in quadrature.

\section{Discussion}

\subsection{X-ray luminosity of the cluster}

With the redshift of the cluster known, the X-ray properties of the
cluster can be estimated, although with of order 85 detected photons,
uncertainties are large.  The temperature of the X-ray emitting gas
was estimated using XSPEC. Photons from all detectors were extracted
from a 34 arcsec radius circle centred on the X-ray centroid.
Following \cite{willis05}, a spectral fit was performed with XSPEC,
using an APEC model \citep{smith01}, abundances from \cite{grevesse99}
and a Galactic neutral hydrogen column of $2.64 \times 10^{20}$
cm$^{-2}$ \citep{dic90}. This resulted in a best fit temperature of
3.8keV, with a 1$\sigma$ lower limit of 1.9keV and an effectively
unconstrained upper limit.

In order to determine a count-rate and hence a luminosity for the
cluster, an aperture over which the rate is to be measured must be
defined. The low number of counts from the object limits the useful
size of the aperture to be around 50-60\arcsec, comparable to those
used in previous X-ray studies of distant clusters. To support this
choice we estimated r500, the radius at which the cluster density is
500 times the critical density at its redshift. Using the best-fit
temperature and following the method in \cite{willis05}, we estimated
this to be 55\arcsec (albeit with considerable uncertainty due to the
uncertainty on the temperature). We used this radius for the subsequent
analysis.  The background-subtracted count rate in the [0.5-2]keV band
within this radius was $6.8 \times 10^{-3}$ cts/s, corresponding to a
flux of $6.2\times 10^{-15}$ erg s$^{-1}$ cm$^{-2}$. Using this to
normalise the spectral fit gives an unabsorbed bolometric luminosity
of $1.1\times 10^{44}$ \ergps with a statistical error of approximately
20 per cent.  Ideally we would apply an aperture correction for flux
beyond 55 \arcsec, but given the limited signal-to-noise this would be
highly uncertain for this object, so we only quote the luminosity
within this aperture. From simulations and experience with other low
count-rate groups and clusters in the XMM-LSS, we found that
statistical errors are usually under-estimates of the true errors on
the luminosities in such cases (\eg see \citealt{d1}), with more
realistic errors twice that of the raw statistical
values. Consequently, we quote an unabsorbed bolometric luminosity of
$1.1\pm 0.7 \times 10^{44}$ \ergps \ and, for comparison to other high
redshift clusters, a rest-frame [0.1-2.4]keV luminosity of $7.7\pm 0.3
\times 10^{43}$ \ergps \ for \clsshort.

Given that only seven cluster galaxies have confirmed redshifts, the
velocity dispersion of the galaxies add little to this, all we can say
here is that the dispersion in these redshifts is consistent with the
derived X-ray luminosity. The above luminosity is low in comparison to
those of other known $z>1$ clusters.  RX~J0848+4453 at $z=1.27$ has
the lowest luminosity of the previously known systems. Using the
values in \cite{stan01} converted to the cosmology used here gives a
[0.1-2.4]keV luminosity of $5.9 \pm 2 \times 10^{43}$ \ergps \
measured in a 35 \arcsec radius aperture. Using an aperture of the
same size for \clsshort \ results in a luminosity of $3 \pm 1.5 \times
10^{43}$ \ergps. Thus \clsshort \ is about as a luminous as
RX~J0848+4453 (and possibly slightly less luminous), given the quoted
uncertainties.

\begin{figure}
\includegraphics[width=8cm]{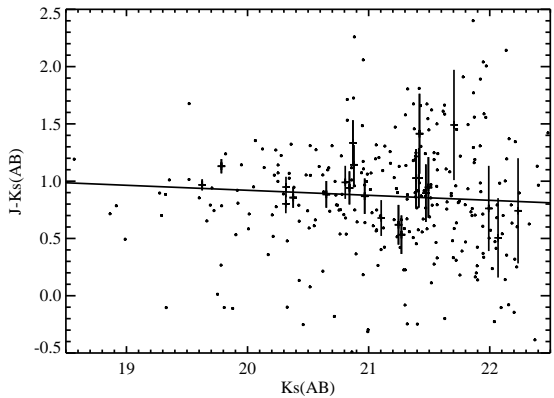}

\caption{$J-Ks$ colour magnitude diagram. Objects from across the
$5\arcmin$ by $5\arcmin$ SOFI field detected in both bands plotted as
small symbols with no error bars. Objects in table 1 and those
detected in the central 250 kpc of the cluster are plotted with error
bars. Overplotted is the best-fit colour-magnitude relation from
Lidman etal., (2004) for RDCS J1252.9-2927 at a similar redshift,
indicating the similarity in the colours of the galaxies in the two
systems.}
\label{cmd1}
\end{figure}

\subsection {Galaxy properties}

Figure \ref{fig:cls} shows the $Ks-$band image of the cluster,
overlayed with the X-ray contours from the XMM data, with the
spectroscopically-confirmed cluster members labelled G1 to G7. Figure
\ref{tci} shows a wider-field pseudo true-colour image of the cluster
made from $z-$band VLT/FORS2 data, $Ks-$band NTT/SOFI data and the
3.6\micron\ Spitzer/IRAC image. The cluster is easily identified in
this image as an obvious overdensity of relatively red sources at the
centre. As noted above, there are fifteen
$19.15<Ks<21.65$ objects identified in a 250 kpc ($30 \arcsec$) diameter
circle centred close to the X-ray peak. Simple visual inspection of
figure \ref{tci} demonstrates that this is an order of magnitude
or more overdensity of such sources in comparison to the surrounding
field.

Figure \ref{tci} also indicates that many of the galaxies in the
central 250 kpc region have similar colours, redder than the average
faint source between $z$ and $3.6\micron$. Figure \ref{cmd1} shows the
$J-Ks$ colour-magnitude diagram of the objects in the SOFI data. The
symbols with error bars denote the data for objects in the central 250
kpc and those listed in table 1. Overlayed is the colour-magnitude
relation found for RDCS1252.9-2927, another z=1.22 cluster, by
\cite{lidman04}. The colours of most of these galaxies are consistent
with this relation, and following the arguments in \cite{lidman04}
indicate that the galaxies have colours consistent with passive
evolution, having formed the bulk of their stellar populations at
higher redshift.

\begin{figure}
\includegraphics[width=8cm]{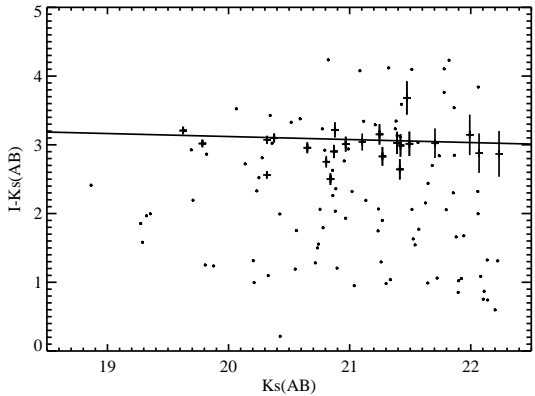}

\caption{$I-Ks$ colour magnitude diagram for the same objects as
figure \ref{cmd1}.  The J-K colour-magnitude relation from
Lidman etal., 2004 has been shifted by 2.2 magnitudes and overplotted on
the data and proves a good fit to the photometry for the galaxies at
the centre of the cluster. }
\label{cmd2}
\end{figure}

\begin{figure}
\includegraphics[width=9cm]{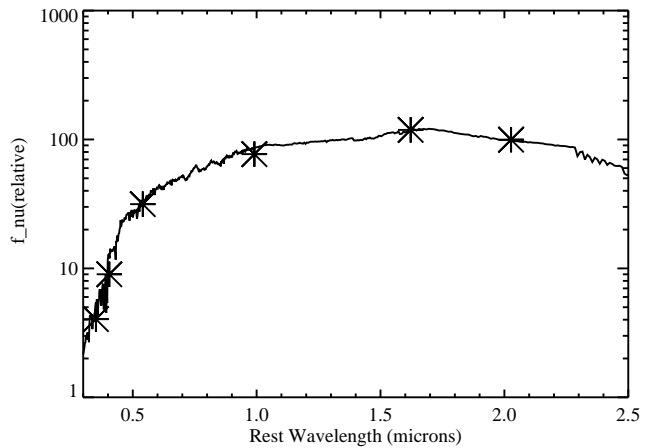}

\caption{Rest-frame spectral energy distribution for the cluster
galaxy G5.  The asterisks denote the $i'$ and $z'$ phrotometry from
CFHTLS, the SOFI $J$ and $Ks$ values and the IRAC $3.6$ and $4.5$
\micron \ photometry. Overplotted is a Bruzual \& Charlot (2003)
instantaneous burst, simple stellar population model with solar
metallicity seen 3 Gyr after the burst. Similar models as young as 1.5
Gyrs can give an acceptable fit to the data assuming some intrinsic
reddening (see text for details).The statistical uncertainties quoted
in table 1 are smaller than the size of the asterisks.}
\label{sed}
\end{figure}

 Figure \ref{cmd2} shows a similar colour-magnitude diagram for
 $I-Ks$. This shows a tight relation between colour and magnitude for
 the confirmed cluster members and those in the central 250
 kpc. Overplotted is the $J-Ks$ colour-magnitude relation from
 \cite{lidman04}, shifted by 2.2 magnitudes (a reasonable shift for a
 z=1.22 elliptical). Most of the cluster members fall on this relation
 within their uncertainties. The tightness of the relation and the
 very red colours again indicates that the majority of these galaxies
 appear to have predominantly passively evolving stellar populations
 dominated by old stars with little significant ongoing star
 formation. For example, in figure \ref{sed} we plot the rest-frame
 spectral energy distribution of G5, the brightest galaxy in the
 central $30\arcsec$ which has colours typical of other galaxies in
 the central region of the cluster. We overplot the best-fit
 instantaneous single burst, solar metallicity simple stellar
 population (SSP) synthesis model from \cite{bc03}. This has an age of
 3~Gyr, implying a redshift of formation of $z=3.2$. However,
 metallicity and reddening can effect the inferred age of the best-fit
 to broad-band SEDs ( \eg \citealt{lidman04} and \citealt{brem02}). We
 attempted to fit younger models reddened by a $\lambda^{-1.3}$ power
 law normalised to $E(B-V)$=0.1. Models as young as 1.5 Gyr gave
 reasonable fits given the uncertainties in the photometry. As found
 by other authors, \eg \citealt{lidman04}, models with more extended,
 but exponentially-declining star formation episodes give comparable
 ages for the bulk of stars in the galaxies.  As noted in section 2.2
 and in figure \ref{fig:spec}, the absorption line spectra of the
 cluster members is well--matched by a \cite{kin96} elliptical
 spectrum, again consistent with a substantial old population of
 stars. Clearly there is some star formation (or AGN) activity in at
 least one of the galaxies, given the presence of the
 [OII]~$\lambda$3727 line in the spectrum of G4. This may be due to
 emission from the galaxy itself, or contamination from another
 cluster member in the spectroscopic slit, possibly the object that is
 included in the IRAC aperture.

Where the colours of the galaxies indicate that they have relatively
old stellar populations, their IRAC and $Ks-$band magnitudes indicate
that these galaxies are already massive at $z=1.22$. If they evolved
passively to $z=0$ they would be approximately 1-1.3 magnitudes
fainter in the rest-frame $K-$band (depending on the age of the
galaxies at $z=1.2$). At $z=1.22$, the 4.5\micron\ band maps almost
directly to the rest-frame $K$. If the system was placed at the
distance and look-back time of Coma, taking into account passive
evolution, galaxy G5 would have $Ks \sim 10.6$ (8.7 in Vega), within a
few tenths of a magnitude of the value for NGC~4889 in
Coma. Similarly, the tenth brightest galaxy in figure \ref{cmd1} is
less than two magnitudes fainter than this, again comparable to the tenth
brightest member in Coma \citep{depropris98}. In common with other
recent studies (\eg \citealt{depropris99,toft04,andreon04}), these
results are consistent with a scenario where the more massive cluster
galaxies are largely in place within a cluster at $z>1$, subsequently
evolving passively with no significant new star formation or
substantial growth by mergers.

How do the properties of these cluster galaxies compare to those in
other known clusters at similar redshifts? \cite{lidman04} and
\citep{blak03} have carried out ground-based near IR and HST-based
optical studies respectively of another z=1.2 cluster, RDCS
J1252.9-2927.  We have already shown that figure \ref{cmd1} is
consistent with figure 2 in \cite{lidman04} -- a similar
colour-magnitude diagram based on SOFI data for that cluster. If we
assume that the three brightest galaxies in the $J-K$ colour-magnitude
sequence of RDCS J1252.9-2927 are the same as the three brightest in
the $i-z$ sequence \citep{blak03}, we infer similar colours for
galaxies in that cluster as for the galaxies in
\clsshort. \cite{stanford97} and \cite{stanford02} found similar
colours for galaxies in the $z=1.27$ cluster RX~J0848+4453 and the
$z=1.16$ cluster around 3C210. All of these authors conclude that such
colours are consistent with a high formation redshift for the stellar
populations of the bulk of the identified bright/massive cluster
galaxies and passive evolution thereafter.

Taking all of the above into account, our current data  is
consistent with most of the identified cluster galaxies at the centre
of the system having dominant stellar populations at least 1.5~Gyr
old, and possibly as old as 3~Gyr, having evolved passively after an
initial burst or a short period of star formation. Their stellar
masses are comparable to those of bright cluster ellipticals at low
redshift, appearing to rule out significant growth by merger for these
galaxies at lower redshifts. That is not to say that that other less
massive galaxies within this cluster are not undergoing significant
merger and star formation activity and can continue to do so, our
$Ks-$band selection naturally selects the most massive galaxies at the
observed epoch.

 The spatial distribution of these galaxies appears more compact than
 in clusters of a similar redshift such as RX~J0848+4453
 (\citealt{stanford97, ros99}) and RDCS J1252.9-2927
 \citep{lidman04}. The surface density of sources brighter than
 $Ks<21.65$ in the central 30 \arcsec of \clsshort \ is $\sim 75$ per
 arcmin$^2$, this compares to $\sim 25$ and $ \sim 45$ per arcmin$^2$
 in the central 70 and 40 arcseconds of these two clusters, estimated
 from the colour magnitude diagrams in \cite{ros99} and
 \cite{lidman04}. The published images of these clusters do not show
 as sharp a drop in surface density of sources beyond 15 \arcsec
 radius. However, it is unclear whether the cluster and its mass
 distribution is truely compact, or whether our current data only
 trace the distribution of the most massive galaxies within it. Deeper
 optical, near-IR and X-ray imaging is clearly required to determine
 the spatial distribution of matter within the cluster, as is further
 optical spectroscopy in order to probe its mass.

\subsection{Detectability of high redshift clusters with combined XMM-LSS and SWIRE data}

The availability of deep CFHTLS optical imaging data over much of the
initial $\sim 5$ deg$^2$ of the XMM-LSS area means that it is
straightforward to identify groups and clusters of galaxies out to
$z\sim 0.8-1$ as extended X-ray sources associated with overdensities
of faint galaxies with similar optical colours (\eg
\citealt{valt04,willis05}). The same data can be used to identify
clusters as galaxy overdensities at higher redshifts, but this becomes
increasingly difficult at $z>1$ and potentially impossible at $z>1.4$.
So long as XMM sources can be reliably identified as extended, one can
have confidence that the absense of an identification in deep optical
data indicates that the source is likely to be a high redshift
cluster. However, given typical $z>1$ clusters with luminosities
around $10^{44}$ \ergps will produce only a few tens of counts in
typical XMM-LSS exposures, confirming that these sources are extended
is challenging. Given the red colours of galaxies in cluster cores, a
sufficiently deep ground-based near-IR or spaced-based Spitzer/IRAC
survey which covers the same area of sky as XMM-LSS can potentially
play the same role at the highest redshifts as the CFHTLS data does at
$z<1$. The clear detection of \cls \ in the SWIRE data is therefore
significant, as this survey covers much of the initial XMM-LSS area. A
similar cluster at even higher redshift would have been detectable as
an excess of galaxies in data as deep as that as SWIRE (\eg the
$z=1.4$ cluster detected by \cite{stan05} would be detectable in IRAC
data of comparable depth to that of  SWIRE). Clusters out to $z\sim
1.5$ discovered by X-ray emission in the XMM-LSS can therefore be
photometrically confirmed by a combination of existing SWIRE/IRAC and
CFHTLS data. A search for such clusters is now underway (Bremer et
al., in prep.)

\section{Conclusions}

We have reported details of the discovery of the $z=1.22$ cluster
\cls, presenting multiband imaging and initial spectroscopy of the
system. We spectroscopically confirm seven galaxies with redshifts of
$z=1.22\pm 0.01$ within an arcminute of the X-ray position. The
cluster appears to have a centrally-condensed galaxy distribution,
with fifteen galaxies with $17.25<Ks<19.75$ within $15 \arcsec$ of the
centre and only a further eight in an annulus between $15\arcsec$ and
$30\arcsec$ from the centre. The spectroscopically-confirmed cluster
members have the colours of passively evolving ellipticals indicating
the bulk of their star formation occurred at least 1.5 Gyr before
$z=1.22$ (\ie at $z>2$). Based on their $Ks$ and IRAC magnitudes, they
have stellar masses comparable with those of massive galaxies in
clusters at low redshift, indicating that massive cluster galaxies may
be in place at $z>1$ and passively evolve at lower redshift with
little significant star formation or growth through mergers. The
straightforward detectability of this cluster in Spitzer/IRAC data
demonstrates that the combination of SWIRE and XMM-LSS datasets allow
for efficient searches for the most distant clusters.

\section*{Acknowledgments}

 XMM is an ESA science mission with instruments and contributions
directly funded by ESA Member States and NASA.  This work was based on
observations made with ESO Telescopes at the La Silla and Paranal
Observatories under programme IDs 70.A-0733 and 074.A-0360 and with
MegaPrime/MegaCam, a joint project of CFHT and CEA/DAPNIA, at the
Canada-France-Hawaii Telescope (CFHT) which is operated by the
National Research Council (NRC) of Canada, the Institut National des
Science de l'Univers of the Centre National de la Recherche
Scientifique (CNRS) of France, and the University of Hawaii. This work
is based in part on data products produced at TERAPIX and the Canadian
Astronomy Data Centre as part of the Canada-France-Hawaii Telescope
Legacy Survey, a collaborative project of NRC and CNRS. Support for
SWIRE, part of the Spitzer Space Telescope Legacy Science Program, was
provided by NASA through an award issued by the Jet Propulsion
Laboratory, California Institute of Technology under NASA contract
1407. This work was also supported by the European Community RTN
Network POE (grant nr. HPRN-CT-2000-00138). MNB acknowledges
Leverhulme Trust funding for the early part of this work. SA
acknowledges financial contribution from contract ASI-INAF
I/023/05/0. We thank Roberto de Propris for useful discussions.

\label{lastpage}

\end{document}